\documentclass[11pt,a4paper]{article}
\usepackage{amsmath,amssymb}
\usepackage[dvipdfmx]{graphicx}
\usepackage{cancel}

\newcommand{\TeV}{\rm{TeV}}
\newcommand{\GeV}{\rm{GeV}}

\newcommand{\lrf}[2]{ \left(\frac{#1}{#2}\right)}

\newcommand{\vev}[1]{\left\langle #1\right\rangle}

\newcommand{\braket}{\vev}

\allowdisplaybreaks

\begin{document}

\begin{titlepage}

\begin{center}

\hfill UT-12-29\\
\hfill IPMU12-0161\\

{\Large \bf 

Scalar Decay into Gravitinos \\
in the Presence of D-term SUSY Breaking

}

\vskip .75in

{\large
Motoi Endo$^{(a,b)}$, 
Koichi Hamaguchi$^{(a,b)}$, 
Takahiro Terada$^{(a)}$
}

\vskip 0.25in

{\em
$^a$Department of Physics, The University of Tokyo,
Tokyo 113-0033, Japan\\
$^b$Kavli Institute for the Physics and Mathematics of the Universe, \\
The University of Tokyo, Kashiwa 277-8568, Japan
}

\end{center}
\vskip .5in

\begin{abstract}
Scalar decay into gravitinos is studied in the presence of D-term supersymmetry (SUSY)  breaking.
The gravitinos produced by the decay of coherent oscillations of scalar fields, such as 
moduli and inflatons, cause cosmological problems.
We show the general formula for the partial decay rate of the scalar into two gravitinos.  
Applying it to a concrete model of D-term SUSY breaking, we find that
 D-term SUSY breaking can suppress gravitino production.

\end{abstract}

\end{titlepage}

%%%%%%%%%%%%%%%%%%%%%%%%%%%%%%%%%%%%%%%%%%%%%%%%%%%%%%%%%%%%%%%%%%%%%%%%%%%%%%%%%%%%%%%%
\section{Introduction}
In models of supergravity (SUGRA)~\cite{Ref:SUGRA,WessBagger}, superweakly interacting massive fields are generally involved, such as the gravitino and modulus fields. They often cause  cosmological problems, such as the moduli or Polonyi problem~\cite{Coughlan:1983ci,Ref:Moduli}, and gravitino problems~\cite{Ref:Gravitino,Kawasaki:2008qe}. If they are generated in an early stage of the universe, they easily spoil big bang nucleosynthesis (BBN) when they decay, or tend to overclose the universe if stable. Hence, their abundances are severely restricted. Among the SUGRA particles, gravitinos are thermally produced in hot plasma. Moreover, it has been pointed out that a large amount of gravitinos are generated via decays of 
coherent oscillations of
heavy scalars, such as moduli and inflatons~\cite{MIGP,Kawasaki:2006gs,Asaka:2006bv,Endo:2007sz}. Therefore, the cosmology constrains the reheating temperature and models of the scalars.

The partial decay rate of a scalar into a pair of gravitinos depends on the structures of the scalar and supersymmetry (SUSY)-breaking sectors. The production rate has been studied in detail by Ref.~\cite{Endo:2006tf} when the SUSY breaking is caused by a vacuum expectation value (VEV) of the F term of the SUSY breaking field. In this case, the goldstino, which is the longitudinal component of the gravitino, is composed of the spinor component of the SUSY-breaking chiral supermultiplet.  
In the literature which studies the gravitino production, D-term contributions to the scalar potential have been discarded. If the SUSY breaking is dominated by a D-term VEV, the goldstino mainly consists of a gaugino, and the above result can significantly change. 
In this paper, direct gravitino production by scalar decay will be studied when the D-term VEV is sizable.

This paper is organized as follows. In Sec.~\ref{sec:general}, the production rate of a pair of gravitinos will be provided, taking the D-term potential into account.  Scalar mixing with the fields in the SUSY-breaking sector will be investigated in the SUGRA framework. In Sec.~\ref{sec:example}, the result will be applied to a D-term SUSY-breaking model, and its cosmological implications will be discussed. Sec.~\ref{sec:conclusion} is devoted to a summary and discussions.

%%%%%%%%%%%%%%%%%%%%%%%%%%%%%%%%%%%%%%%%%%%%%%%%%%%%%%%%%%%%%%%%%%%%%%%%%%%%%%%%%%%%%%%%
\section{Direct Gravitino Production Rate}\label{sec:general}

In this section, we evaluate the partial decay rate of a scalar field such as the modulus or inflaton field into a pair of gravitinos.  It is derived at the leading order of Planck-suppressed interactions  in the framework of SUGRA.
Both the D-term and F-term SUSY-breaking contributions are taken into account,
and sources of the D-term contributions are clarified.

%%%%%%%%%%%%%%%%%%%%%%
\subsection{Gravitino Production Rate in the Mass-Eigenstate Basis}

The SUSY-breaking sector is supposed to have nonzero VEVs of the D-term potentials for gauge symmetries which are not included in the Standard Model (SM) gauge groups. The gauge symmetries can be $U(1)$ and/or non-Abelian symmetries. 
If the symmetry is $U(1)$, we assume that there is no genuine Fayet-Iliopoulos term, since it is difficult to embed the Fayet-Iliopoulos term in SUGRA~\cite{FI-SUGRA}.
The VEVs of the D-term potentials are considered to be dynamically generated by the fields in the SUSY-breaking sector, $z^i$, which are charged under the gauge symmetries. See Sec.~\ref{sec:example} for an example of such a model.

In gravitino production, the relevant fields are the scalar field, $\phi$, the SUSY-breaking fields, $z^i$, and the gravitino. 
It is assumed that the scalar $\phi$ is much heavier than the gravitino due to the SUSY-invariant mass.\footnote{This is the case for the inflaton in many inflation models. In the case of modulus, if the modulus mass is $m_\phi < {\cal O}(10 \TeV)$, its decay occurs after BBN starts and spoils the success of the standard cosmology. Thus, we assume $m_\phi \gg 10 \TeV \gtrsim m_{3/2}$.}
The scalar is also assumed to be a singlet under the extra gauge symmetry, 
and couples with $z^i$ only through the terms suppressed by powers of the (reduced) Planck scale, $M_P = 2.4 \times 10^{18} \GeV$, which generally exist in the SUGRA Lagrangian.\footnote{Otherwise, $\phi$ decays into the SUSY-breaking sector much faster, which worsens the cosmological gravitino problem.}

The superpotential, K\"ahler potential and gauge kinetic function are represented as\footnote{In this paper, we follow the conventions of Ref.~\cite{WessBagger}. 
Derivatives with respect to fields are denoted by subscripts, e.g., $G_{\phi}=\partial G/\partial \phi$.
Also, we omit symbols of VEV, $\vev{\cdots}$, if not otherwise specified.
}
\begin{align}
K &= K(\phi ,\bar{\phi})+K(z^{i} ,\bar{z}^{i}) 
+ \sum_{n\ge 1}\frac{1}{M_P^n}K_{\rm mix}^{(n)}(\phi, \bar{\phi}, z^i, \bar{z}^i),  \nonumber\\ 
W &= W(\phi) + W(z^i) + \sum_{n\ge 1}\frac{1}{M_P^n}W_{\rm mix}^{(n)}(\phi, z^i), \nonumber\\
h_{AB}&=\left( 1+\frac{\theta g^{2}}{8 \pi i} \right) \delta_{AB} 
+\sum_{n\ge 1}\frac{1}{M_{P}^{n}}h_{AB}^{(n)}(\phi, z^{i}),
\label{eq:model-base}
\end{align}
where $K_{\rm mix}^{(n)}$ and $W_{\rm mix}^{(n)}$ are the interaction terms between $\phi$ and $z^i$ whose Planck suppressions are displayed explicitly. The first term in the right-hand side of $h_{AB}$ corresponds to the gauge kinetic and $\theta$ terms, and  $g$ is the gauge coupling constant of the extra gauge symmetry.
Let us call the basis in Eq.~\eqref{eq:model-base} the ``model basis'' in this paper. 
Note that the VEVs of $\phi$ and $z^i$ can induce kinetic mixings through higher-dimensional terms in the K\"ahler potential. In the following, we assume that those VEVs are much smaller than the Planck scale. Otherwise, the following discussion applies after the fields are shifted to absorb the large VEVs.

The decay rate is evaluated in the mass-eigenstate basis. The fields in the mass-eigenstate basis, $X^{a}$, are related to those in the model basis, $x^{\alpha}$, as
\begin{eqnarray}
X^{a}=A^{X^{a}}{}_{x^{\alpha}} \delta x^{\alpha},
\end{eqnarray}
where $\delta  x^{\alpha}= x^{\alpha}-\braket{x^{\alpha}}$ is the fluctuation around the VEV, and $A$ is a matrix to diagonalize the mass matrix and to canonicalize the kinetic terms at the potential minimum. The model basis fields, $x^{\alpha}$, consist of $x^{\alpha}=\varphi^{\alpha}$ and its Hermitian conjugate $\bar{\varphi}^{\alpha}$, with $\varphi^{\alpha}=\phi$ and $z^{i}$. 
On the other hand, the mass eigenstates are generally represented by
real fields, $X^{a}=\Phi_{R}, \Phi_{I}, Z_{R}^{i}$, and $Z_{I}^{i}$, which primarily consist of the 
real and imaginary parts of $\phi$ and $z^i$, respectively.
Hence, $(A^{X^{a}}{}_{\varphi^{\alpha}})^* = A^{X^{a}}{}_{\bar{\varphi}^{\alpha}}$ is satisfied.
Since the interactions between $\phi$ and $z^i$ are given by higher-dimensional operators,
$A^{\Phi_{R,I}}{}_{z^{i}}, A^{\Phi_{R,I}}{}_{\bar{z}^{i}}, A^{Z^{i}_{R,I}}{}_{\phi}$, and $A^{Z^{i}_{R,I}}{}_{\bar{\phi}}$ are suppressed by the Planck scale.
The matrix $A$ will be evaluated later.

 The Lagrangian terms which are relevant for the tree-level decay of the scalar into a pair of gravitinos are found to be~\cite{WessBagger}
 \begin{align}
\mathcal{L}=& \frac{1}{8i}\epsilon ^{\mu \nu \rho \sigma}( G_{\varphi^{\alpha}} \partial_{\rho} \delta \varphi^{\alpha} - G_{\bar{\varphi}^{\alpha}}\partial_{\rho}\delta \bar{\varphi}^{\alpha} )\bar{\psi}_{\mu}\gamma_{\nu}\psi_{\sigma} \nonumber \\
& +\frac{1}{4}m_{3/2} ( G_{\varphi^{\alpha}} \delta \varphi^{\alpha} + G_{\bar{\varphi}^{\alpha}}\delta \bar{\varphi}^{\alpha} ) \bar{\psi}_{\mu} \gamma ^{\mu \nu} \psi_{\nu}, \label{Lm22g}
\end{align}
where $\varphi^{\alpha}$ stands for $\phi$ and $z^{i}$ in the model basis, while $\psi_{\mu}$ is the gravitino, $m_{3/2}$ the gravitino mass, and $G=K+\ln |W|^{2} $ the total K\"{a}hler potential.
Here and hereafter, the Planck unit, $M_{P} = 1$, is used if not otherwise specified.
This Lagrangian is the same as that in the F-term SUSY-breaking case~\cite{MIGP}.
In terms of the mass-eigenstate basis, the above Lagrangian becomes 
\begin{align}
\mathcal{L}=& \frac{1}{8i}\epsilon ^{\mu \nu \rho \sigma} \mathcal{G}^{\text{(1)}}_{\Phi_{R}} \partial_{\rho} \Phi_{R} \bar{\psi}_{\mu}\gamma_{\nu}\psi_{\sigma}  +\frac{1}{4}m_{3/2}  \mathcal{G}^{\text{(2)}}_{\Phi_{R}}  \Phi_{R} \bar{\psi}_{\mu} \gamma ^{\mu \nu} \psi_{\nu} + (R \rightarrow I) + \dots,
\label{eq:g12RI}
\end{align}
where the coefficients are
\begin{align}
\mathcal{G}^{\text{(1)}}_{\Phi_{R,I}} =2i\, \Im ( G_{\varphi^{\alpha}}(A^{-1})^{\varphi^{\alpha}}{}_{\Phi_{R,I}} ),~~~
\mathcal{G}^{\text{(2)}}_{\Phi_{R,I}} =2\, \Re ( G_{\varphi^{\alpha}}(A^{-1})^{\varphi^{\alpha}}{}_{\Phi_{R,I}} ),
\end{align}
where $A^{-1}$ is the inverse matrix of $A$, and 
 $\Re(\cdots)$ and $\Im(\cdots)$ represent the real and imaginary parts, respectively.
The omitted terms in Eq.~(\ref{eq:g12RI}) include $Z_{R,I}^{i}$, which are irrelevant for the gravitino production by the $\Phi_{R,I}$ decay. Note that the gravitino production rate will be derived at the tree level in this section. The rate can receive radiative corrections, which will be mentioned later.

From the above interactions, the partial decay rate of the scalar into a pair of gravitinos is evaluated as~\cite{MIGP}
\begin{align}
\Gamma(\Phi_{R,I}\to \psi_{3/2}\psi_{3/2}) =
\frac{|\mathcal{G}^{\text{(eff)}}_{\Phi_{R,I}}|^{2}m_{\Phi_{R,I}}^{5}}{288 \pi m^{2}_{3/2 }}, 
\label{gravitinorate}
\end{align}
where we have used $m_{\Phi_{R,I}}\gg m_{3/2}$.
The effective coupling constants are defined as
\begin{eqnarray}
\left | \mathcal{G}^{\text{(eff)}}_{\Phi_{R,I}} \right |^{2}
=\frac{1}{2}\left( \left| \mathcal{G}^{\text{(1)}}_{\Phi_{R,I}} \right|^{2} + \left| \mathcal{G}^{\text{(2)}}_{\Phi_{R,I}} \right|^{2} \right) 
=2\left| \sum_{\varphi^{\alpha}=\phi,z^i} G_{\varphi^{\alpha}} (A^{-1}) ^{\varphi^{\alpha}}{}_{ \Phi_{R,I}} \right|^{2}.\label{Geffective}
\end{eqnarray}
The rate is apparently the same as that in the F-term SUSY-breaking models~\cite{MIGP} because the relevant Lagrangian [Eq.~\eqref{Lm22g}] is the same. The effective coupling constants [Eq.~\eqref{Geffective}] are governed by $G_{\varphi^{\alpha}}$, which is related to the F term as 
\begin{align}
F^i = -e^{G/2} g^{i \bar j} G_{\bar j},
\end{align}
where $g^{\bar{i}j}$ is the inverse of the K\"{a}hler metric, $g_{i\bar{j}}=\vev{G_{i\bar{j}}}$.
However, this does not mean that the D term is unimportant.
As we will see, the magnitudes of F terms are controlled by vacuum conditions,
which are affected by the D-term potential.
Moreover, the D term contributes to the scalar mass matrix, 
which determines the mixing of scalar fields with the SUSY-breaking fields.
These D-term contributions to the effective coupling constants are studied in the following subsection.

%%%%%%%%%%%%%%%%%%%%%%
\subsection{Effective Coupling Constants}
\label{sec:EffCoupling}

In this subsection, the effective coupling constants of gravitino pair production, $\mathcal{G}^{\text{(eff)}}_{\Phi_{R,I}}$, are evaluated in the model basis within the SUGRA framework. Since we are interested in the cosmological applications of gravitino production, it is sufficient to evaluate them at the leading order with respect to the Planck-suppressed couplings. Higher-order corrections are safely neglected. 
The interactions are classified by powers of the inverse of the Planck scale. 
At the zeroth order, i.e., in the global SUSY limit, $\phi$ is secluded from the SUSY-breaking sector by the assumptions. On the other hand, SUGRA corrections which include Planck-suppressed interactions belong to higher orders of the perturbation. Turning on the corrections, the two sectors communicate with each other, and  gravitino production occurs.

In Eq.~\eqref{Geffective}, it is sufficient to evaluate $G_{z^{i}}$ and $(A^{-1})^{\phi}{}_{ \Phi_{R,I}}$ at the zeroth order, because $G_{\phi}$ and $(A^{-1})^{z^{i}}{}_{ \Phi_{R,I}}$ start from the first order of the perturbation. Let us start from $(A^{-1})^{\phi}{}_{ \Phi_{R,I}}$.
In the global SUSY limit, the mass eigenstates in the $\phi$ sector are simply given by $\Phi_{R}=\sqrt{g_{\phi\bar{\phi}}/2}(\delta\phi+\delta\bar{\phi})$ and $\Phi_{I}=-i\sqrt{g_{\phi\bar{\phi}}/2}(\delta\phi-\delta\bar{\phi})$, with their masses
\begin{align}
m^{2}_{\Phi_R}=\frac{1}{g_{\phi \bar{\phi}}} \left( V^{\rm (g)}_{\phi \bar{\phi}} + V^{\rm (g)}_{\phi \phi} \right) ,~~~~~
m^{2}_{\Phi_I}=\frac{1}{g_{\phi \bar{\phi}}} \left( V^{\rm (g)}_{\phi \bar{\phi}} - V^{\rm (g)}_{\phi \phi} \right) ,
\end{align}
where $V^{\rm (g)}$ denotes the scalar potential in the global SUSY limit, and 
\begin{eqnarray}
V_{x^\alpha x^\beta} = \frac{\partial^2 V}{\partial x^\alpha \partial x^\beta}
\end{eqnarray}
is the second derivative of the potential in the model basis. Here, we assume $V^{\rm (g)}_{\phi \phi} = V^{\rm (g)}_{\bar{\phi}\bar{\phi}}$ for simplicity. It is straightforward to include the phase.
Thus, the mixings of the scalar $\phi$ itself become $(A^{-1})^{\phi}{}_{ \Phi_{R}}=1/\sqrt{2g_{\phi\bar{\phi}}}$ and $(A^{-1})^{\phi}{}_{ \Phi_{I}}=i/\sqrt{2g_{\phi\bar{\phi}}}$ at the zeroth order. 
The effective coupling constants [Eq.~\eqref{Geffective}] are approximated to be
\begin{align}
\left| \mathcal{G}^{\text{(eff)}}_{\Phi_{R}}\right|^{2}=2\left|\frac{1}{\sqrt{2g_{\phi \bar{\phi}}}} G_{\phi} + G_{z^{i}} (A^{-1})^{z^{i}}{}_{\Phi_{R}} \right|^{2},  \label{GeffR0} \\
\left| \mathcal{G}^{\text{(eff)}}_{\Phi_{I}}\right|^{2}=2\left|\frac{i}{\sqrt{2g_{\phi \bar{\phi}}}} G_{\phi} + G_{z^{i}} (A^{-1})^{z^{i}}{}_{\Phi_{I}} \right|^{2}.  \label{GeffI0}
\end{align}
Next, the F terms of the SUSY-breaking fields, $G_{z^{i}}\simeq W_{z^i}/W$, are evaluated at the zeroth order, namely by using the field VEVs in the global SUSY limit, as discussed above. This depends on the model, and we will demonstrate it in a D-term SUSY-breaking model in the next section. Although the SUSY-breaking sector may be involved, this procedure is straightforward. 

The remaining task is to evaluate $G_{\phi}$ and $(A^{-1})^{z^{i}}{}_{\Phi_{R,I}}$ at the first order. 
They include contributions from the D-term potential, which is represented by the Killing potential, $D_{A}$. 
By the gauge invariance of the action, the Killing potential satisfies
\begin{align}
D_A = i X_A^i G_i = i X_A^i K_i,
\label{eq:D-term}
\end{align}
where $X_A^i =-ig^{i\bar{j}}D_{A \bar{j}}$ is the holomorphic Killing vector, by which the gauge transformation of a chiral superfield, $\Phi^i$, is defined as $\delta \Phi^i = \Lambda^A X_A^i (\Phi^j)$ with $\Lambda^A$ being the gauge transformation parameters. The second equality in Eq.~\eqref{eq:D-term} is derived by the gauge invariance of the superpotential, $\delta W = W_i \delta \Phi^i = 0$. 

The F term of $\phi$ is evaluated at the minimum of the scalar potential in SUGRA. The  conditions of the vanishing cosmological constant and the potential minimization are\begin{align}
&V = \frac{1}{2}g^{2} D^{A}D_{A} + e^{G}(G^{i}G_{i}-3) = 0, \label{V} \\
&V_{i}= g^{2} \left(-\frac{1}{2}h^{R}_{ABi}D^{A}D^{B}+D^{A}D_{Ai} \right)+e^{G}\left( G_{i}G^{j}G_{j}-2G_{i}+G^{j}\nabla _{i} G_{j} \right)=0, \label{Vi} 
\end{align}
where $V$ is the scalar potential in SUGRA,  $h^{R}_{AB}\, (h_{R}^{AB})$ is (the inverse of) the real part of the gauge kinetic function,
and $\nabla_{i}G_{j}=G_{ij}-\Gamma^{k}_{ij}G_{k}$ is the covariant derivative with $\Gamma^{k}_{ij}=g^{\bar{l}k}G_{ij\bar{l}}$.
From Eq.~(\ref{Vi}), $G_{\phi}$ is obtained as
\begin{align}
g^{\phi\bar{\phi}} G_{\phi} \simeq 
\frac{1}{\nabla_{\bar{\phi}}G_{\bar{\phi}}} \left[ -g^{\bar{z}^{i}z^{j}}G_{z^{j}} \nabla_{\bar{\phi}}G_{\bar{z}^{i}}  +\frac{g^{2}}{m_{3/2}^{2}} \left( \frac{1}{2}h^{R}_{AB\bar{\phi}}D^{A}D^{B} - D^{A}D_{A\bar{\phi}} \right) \right]
- g^{z^{i}\bar{\phi}} G_{z^{i}}, 
\label{Gphi}
\end{align}
where the gravitino mass is from $m_{3/2} = \braket{e^{G/2}}$. 
In this expression, we have used $|\nabla_{\phi}G_{\phi}| = W_{\phi\phi}/W + \cdots \sim m_\phi/m_{3/2} \gg 1$ and  $G_i \lesssim {\mathcal O}(1)$ from Eq.~\eqref{V}.
It is found that $G_{\phi}$ vanishes in the global SUSY limit, when $\phi$ is secluded from the SUSY-breaking sector in the limit and is a singlet under the extra gauge symmetry.

Next, let us evaluate the mixing matrix $(A^{-1})^{z^{i}}{}_{\Phi_{R,I}}$. There are two sources of mixing between $\phi$ and $z^{i}$. The first one is from kinetic mixing, $g_{z^{i}\bar{\phi}}$, which is induced by higher-dimensional operators with the field VEVs. The kinetic terms are canonicalized by redefining the fields as
\begin{align}
\phi'=\sqrt{g_{\phi\bar{\phi}}}\;\delta\phi+\frac{g_{z^{i}\bar{\phi}} }{\sqrt{g_{\phi\bar{\phi}}} }\delta z^{i},~~~
z'^{i}= (C^{-1})^{i}{}_{j}\delta z^{j},
\label{non-unitary}
\end{align}
at the leading order of $g_{z^{i}\bar{\phi}}$,
where $C$ is a matrix that canonicalizes the kinetic term of the SUSY-breaking sector,
\begin{align}
C^{\dag}{}_{i}{}^{j} g_{\bar{z}^{j}z^{k}} C^{k}{}_{l}=\delta_{il}.
\end{align}
Here, it is sufficient to evaluate $g_{z^{i}\bar{\phi}}$ and $g_{\phi\bar{\phi}}$ 
by using the field VEVs in the global SUSY, when the first-order perturbation is considered. 
The second source of the mixing comes from the mass term. By canonicalizing the kinetic terms, the mixings in the mass matrix become
\begin{align}
V_{\phi' \bar{z}^{\prime i}} &= 
\frac{1}{\sqrt{g_{\phi\bar{\phi}}}} (C^\dagger)_i{}^j
\left[ V_{\phi \bar{z}^{j}} - \frac{g_{\phi \bar{z}^{j}}}{g_{\phi\bar{\phi}}} V^{\rm (g)}_{\phi\bar{\phi}} \right] 
\equiv \frac{1}{\sqrt{g_{\phi\bar{\phi}}}} (C^\dagger)_i{}^j\, \tilde{V}_{\phi \bar{z}^{j}}, 
\label{eq:mass-matrix-1} \\
V_{\phi' z^{\prime i}} &= 
\frac{1}{\sqrt{g_{\phi\bar{\phi}}}} C^j{}_i
\left[ V_{\phi z^{j}} - \frac{g_{z^{j}\bar{\phi}}}{g_{\phi\bar{\phi}}} V^{\rm (g)}_{\phi \phi} \right]
\equiv \frac{1}{\sqrt{g_{\phi\bar{\phi}}}} C^j{}_i\, \tilde{V}_{\phi {z}^{j}},
\label{eq:mass-matrix-2}
\end{align}
at the leading order of $g_{z^{i}\bar{\phi}}$. 
Explicit forms of $V_{\phi z^{i}}$ and $V_{\phi \bar{z}^{j}}$ will be given later.
Since the mixing terms are small, the mass matrix is diagonalized by means of the perturbation theory. At the leading order of the perturbation, $(A^{-1})^{z^{i}}{}_{\Phi_{R,I}}$ is obtained by combining the canonicalization and the diagonalization as
\begin{align}
(A^{-1})^{z^{i}}{}_{ \Phi_{R}}&= \frac{1}{\sqrt{2g_{\phi\bar{\phi}}}}
\left[ \left( m^{2}_{\Phi_{R}} g_{z} -m_z^{2} \right)^{-1} \right]^{z^{i}\tilde{z}^{j}} \left( \tilde{V}_{\tilde{z}^{j} \phi}+ \tilde{V}_{\tilde{z}^{j} \bar{\phi}} \right), \\
(A^{-1})^{z^{i}}{}_{ \Phi_{I}}&= \frac{i}{\sqrt{2g_{\phi\bar{\phi}}}}
\left[  \left( m^{2}_{\Phi_{I}} g_{z} -m_z^{2} \right)^{-1}\right]^{z^{i}\tilde{z}^{j}} \left( \tilde{V}_{\tilde{z}^{j} \phi}- \tilde{V}_{\tilde{z}^{j} \bar{\phi}} \right),
\end{align}
where the indices $\phi$ and $z^i$ are not in the primed basis of Eq.~\eqref{non-unitary}, but in the model one. 
Here, $\left( m^{2}_{\Phi} g_{z} - m_z^{2} \right)^{-1}$ is an inverse matrix, and the index $\tilde{z}^{j}$ runs over both $z^{j}$ and $\bar{z}^{j}$. The matrices, $m_z^{2}$ and $g_z$, are defined in the model basis as 
\begin{align}
m_z^2 = &\begin{pmatrix}
V^{\rm (g)}_{z^i\bar z^j} & V^{\rm (g)}_{z^i z^j} \\
V^{\rm (g)}_{\bar z^i \bar z^j} & V^{\rm (g)}_{\bar z^i z^j} 
\end{pmatrix},
~~~~~
g_{z}=
\begin{pmatrix}
g_{z^i\bar z^j} & 0 \\
0 & g_{\bar z^i z^j}
\end{pmatrix}.\label{zmatrices}
\end{align}
Since the mixing 
$\tilde{V}_{\tilde{z}^{j} \phi}\pm \tilde{V}_{\tilde{z}^{j} \bar{\phi}}$
belongs to at least the first order of the perturbation, 
the inverse matrix is evaluated in the global SUSY.

The mixing terms $V_{\phi \bar{z}^{i}}$ and $V_{\phi z^{i}}$ are obtained from the scalar potential of SUGRA. 
By using Eqs.~\eqref{V} and  \eqref{Vi},
the mass matrices are derived as
\begin{align}
\label{hmass} 
V_{i \bar{j}} &= 
\mathrm{e}^{G} 
\left( \nabla _{i} G_{k} \nabla _{\bar{j}} G^{k} - R_{i \bar{j} k \bar{l} } G^{k}G^{\bar{l}} + g_{i \bar{j}}  \right) \\
& + g^{2} \left(  \frac{1}{2} \left( G_{i}G_{\bar{j}}  - g_{i \bar{j}} \right) D^{A}D_{A} -G_{i}D^{A}D_{A \bar{j}} - G_{\bar{j}}D^{A}D_{Ai} +\frac{1}{2} \left( h^{R}_{ABi}G_{\bar{j}} + h^{R}_{AB \bar{j}}G_{i} \right) D^{A}D^{B}  \right) \nonumber \\
&  + g^{2} \left( h_{R}^{AB}D_{Ai}D_{B\bar{j}} + h_{R}^{AB}{}_{i}D_{A}D_{B\bar{j}}+ h_{R}^{AB}{}_{\bar{j}}D_{A}D_{Bi} +D^{A}D_{Ai\bar{j}} + h^{R}_{ACi}h_{R}^{CD}h^{R}_{DB\bar{j}} D^{A}D^{B}   \right), \nonumber \\
\label{ahmass}
V_{ij} &= 
\mathrm{e}^{G}
\left( 2\nabla _{i} G_{j} + G^{k}\nabla _{i} \nabla _{j} G_{k} \right) \nonumber \\
& + g^{2} \left( \frac{1}{2} \left( G_{i}G_{j} -\nabla _{i}G_{j} \right)D^{A}D_{A} -G_{i}D^{A}D_{Aj} - G_{j}D^{A}D_{Ai}  \right. \\
& ~~~~~~~~
\left. +\frac{1}{2} \left( h^{R}_{ABi}G_{j} + h^{R}_{ABj}G_{i}  \right)D^{A}D^{B}  +\Gamma ^{k}_{ij} \left( -D^{A}D_{Ak}+\frac{1}{2}h^{R}_{ABk}D^{A}D^{B}  \right)  \right) \nonumber \\
 & +g^{2} \left( h_{R}^{AB}D_{A}D_{Bij}+ h_{R}^{AB}D_{Ai}D_{Bj} 
 + h_{R}^{AB}{}_iD_{A}D_{Bj} 
 + h_{R}^{AB}{}_jD_{A}D_{Bi} 
 + \frac{1}{2}h_{R}^{AB}{}_{ij}D_{A}D_{B} \right), \nonumber
\end{align}
with
\begin{align}
\nabla_{i}\nabla_{j}G_{k}
&= (\nabla_{j}G_{k})_{i} -\Gamma^{l}_{ij}\nabla_{l}G_{k}-\Gamma^{l}_{ik}\nabla_{j}G_{l} \nonumber \\
&= G_{ijk}-G_{ijk\bar{l}}G^{\bar{l}}-3\Gamma^{l}_{(ij}G_{k)l}+3\Gamma^{l}_{(ij}\Gamma^{m}_{k)l}G_{m},
\end{align}
where the indices in a parenthesis are totally symmetrized.
In each of Eqs. \eqref{hmass} and (\ref{ahmass}), the first parenthesis is induced by the F-term VEVs, while the others are finite when the D term contributes. 
At the first order, the mixing terms of the mass matrices which are (potentially) relevant for the cosmology are obtained as
\begin{align}
V_{\phi \bar{z}^{i}} &\simeq 
e^{G} \left( g^{\phi\bar{\phi}}\nabla_{\phi}G_{\phi}\nabla_{\bar{z}^{i}}G_{\bar{\phi}} 
+ g^{z^{j}\bar{z}^{k}}\nabla_{\phi}G_{z^{j}}\nabla_{\bar{z}^{k}}G_{\bar{z}^{i}}
\right. \nonumber\\ &~~~~~~~~~~\left.
+ g^{\phi \bar{z}^{j}}\nabla_{\phi}G_{\phi}\nabla_{\bar{z}^{i}}G_{\bar{z}^{j}} 
- R_{\phi \bar{z}^i z^j \bar{z}^k} g^{z^{j}\bar{z}^{l}}g^{\bar{z}^{k}z^{m}}G_{\bar{z}^l} G_{z^m} \right) 
\nonumber \\ &~~~
+ g^{2} \left( h_{R}^{AB}D_{A\phi}D_{B\bar{z}^{i}}+h_{R}^{AB}{}_{\phi}D_{A}D_{B\bar{z}^{i}}+D^{A}D_{A\phi\bar{z}^{i}}\right), \label{Mphizbar} \\
V_{\phi z^{i}} &\simeq
- e^{G} \left( G_{\bar{\phi}z^{i}z^{j}}G_{\phi\phi}+2g^{z^{k}\bar{z}^{l}}G_{\bar{z}^{l}\phi (z^{i}}G_{z^{j})z^{k}} \right) g^{z^{j}\bar{z}^{m}} G_{\bar{z}^{m}} 
\nonumber \\ &~~~
+ g^{2} \left( -g^{z^{j}\bar{z}^{k}}G_{\phi z^{i}\bar{z}^{k}}D^{A}D_{A z^{j}} 
+ h_{R}^{AB}D_{A}D_{B\phi z^{i}} 
+ h_{R}^{AB}D_{A\phi}D_{B z^{i}} 
+ h_{R}^{AB}{}_{\phi}D_{A}D_{B z^{i}} \right). \label{Mphiz} 
\end{align}
Here, we have used $|G_\phi| \ll 1$ and kept only the terms which can be enhanced by
the following means: (i) the modulus/inflaton is much heavier than the gravitino due to its SUSY mass, $|\nabla_{\phi}G_{\phi}| \gg 1$, (ii) in the D-term SUSY-breaking models, the fields in the SUSY-breaking sector, $z^i$, can have a large SUSY-invariant mass, $|\nabla_{z^i}G_{z^j}| \gg 1$, and (iii) derivatives of $D_A$ with respect to $z^i$ can be enhanced, since the VEVs of $z^i$ are much smaller than $M_P$. 

In summary, the effective coupling constants are represented in the model basis as
\begin{align}
\left| \mathcal{G}^{\text{(eff)}}_{\Phi_{R}}\right|^{2}=
\frac{1}{g_{\phi\bar{\phi}}} \left| G_{\phi}+ G_{z^{i}} \left[ \left( m^{2}_{\phi_{R}} g_{z} - m_z^{2} \right)^{-1} \right]^{z^{i}\tilde{z}^{j}} \left( \tilde{V}_{\tilde{z}^{j} \phi}+ \tilde{V}_{\tilde{z}^{j} \bar{\phi}} \right) \right|^{2},  \label{GeffR} \\
\left| \mathcal{G}^{\text{(eff)}}_{\Phi_{I}}\right|^{2}=
\frac{1}{g_{\phi\bar{\phi}}}\left| G_{\phi}+ G_{z^{i}} \left[\left( m^{2}_{\phi_{I}} g_{z} - m_z^{2} \right)^{-1}\right]^{z^{i}\tilde{z}^{j}} \left( \tilde{V}_{\tilde{z}^{j} \phi}- \tilde{V}_{\tilde{z}^{j} \bar{\phi}} \right) \right|^{2}, \label{GeffI}
\end{align}
where $G_\phi$ is found in Eq.~\eqref{Gphi}, $m^{2}_{z}$ and $g_{z}$ are in Eq.~\eqref{zmatrices}, and the mixing terms are in Eqs.~\eqref{eq:mass-matrix-1} and \eqref{eq:mass-matrix-2} with Eqs.~\eqref{Mphizbar} and \eqref{Mphiz}. They are independent of the matrix $C$ in Eq.~\eqref{non-unitary}. The direct gravitino production rate is evaluated by substituting the above coupling constants into Eq.~\eqref{gravitinorate}. 

The D-term contributions are found in the terms proportional to $g^2$ in Eqs.~\eqref{Gphi}, \eqref{Mphizbar}, and \eqref{Mphiz}. They are classified into two groups: i) contributions of $\phi$ to the gauge kinetic term such as $h_{AB} \sim \phi/M_P$, and ii) contributions to the D-term potential such as $G_{\phi z^{i}\bar{z}^{j}}g^{\bar{z}^{j} z^{k}}D_{Az^{k}}$ and $D_{A\phi}=ig_{\phi \bar{z}^{i}}X^{\bar{z}^{i}}_{A}=g_{\phi \bar{z}^{i}}g^{\bar{z}^{i} z^{j}}D_{A z^{j}}$, due to an effective charge which is induced by mixings with the SUSY-breaking fields from higher-dimensional operators.
See Eq.~\eqref{eq:D-term} for the expression of the D term.
 Thus, all the D-term contributions vanish if the following two conditions are satisfied:
\begin{enumerate}
\item[(I)] $\phi$ does not appear in the gauge kinetic function of the extra gauge symmetry, $h_{AB\phi}=0$.
\item[(II)] $\phi$ does not have the specific K\"{a}hler mixings with the fields in the SUSY-breaking sector which are charged under the extra gauge symmetry, $g_{\phi\bar{z}^i}=G_{\phi z^{i}\bar{z}^{j}}=0$.
\end{enumerate}
If these conditions are satisfied, the gravitino production is only from the F-term contributions. (See Appendix~\ref{app:F-term} for a revision of the F-term contribution.)  It is found that the rate decreases as the D-term VEV increases. This is because 
the VEV of the F-term potential satisfies the vanishing cosmological constant condition [Eq.~\eqref{V}]. If the D-term potential dominates the SUSY breaking, the F-term VEVs become suppressed. Then the cosmological problem of the direct gravitino production can be relaxed. 

Finally, let us comment on radiative corrections to pair gravitino production. The scalar field couples to a pair of gauginos via matter loops analogously to the anomaly, where the scalar interactions with the matter come from gravitational effects or higher-dimensional operators~\cite{Endo:2007ih,Endo:2007sz}. Since the gaugino in the SUSY-breaking sector is a main component of the goldstino, the anomaly-induced decay is considered to contribute to the pair gravitino production. This decay induced by the gravitational anomalies works when the matter fields in the loop are lighter than the scalar. 

When the scalar $\phi$ is heavier than the fields in the SUSY-breaking sector, the latter fields other than the goldstino are also produced by the scalar decay through SUGRA interactions~\cite{Endo:2006qk,Endo:2007sz}. The decay products in the SUSY-breaking sector are considered to decay into lighter fields including the gravitino. This increases the abundance of the gravitino, and thus worsens the cosmology. 

%%%%%%%%%%%%%%%%%%%%%%%%%%%%%%%%%%%%%%%%%%%%%%%%%%%%%%%%%%%%%%%%%%%%%%%%%%%%%%%%%%%%%%%%
\section{Example}\label{sec:example}

In this section, we apply the formulas in the previous section to a model of the D-term SUSY breaking.
A fraction of the D-term SUSY breaking is defined as
\begin{align}
\delta=\frac{V_{D}}{V_{F}+V_{D}}, \label{delta}
\end{align}
where the F-term SUSY breaking is represented by the F-term potential, $V_{F}=e^{G}G^{i}G_{i}$, and the D-term SUSY breaking is represented by the D-term potential, $V_{D}=(g^{2}/2)D^{A}D_{A}$.
In the global SUSY, the F-term potential leads to $V_{F}=g^{i\bar{j}}\overline{W}_{\bar{j}}W_{i}$.
The denominator of the right-hand side of Eq.~\eqref{delta} is related to the gravitino mass through the cosmological constant condition [Eq.~\eqref{V}], i.e.,
\begin{align}
 V_F + V_D = g^{i\bar{j}}\overline{W}_{\bar{j}}W_{i} + \frac{1}{2}g^{2} D^{A}D_{A}=3 m_{3/2}^2 . \label{V2}
\end{align}
The fraction $\delta$ takes a value of $0\leq \delta <1$, where a pure D-term SUSY breaking, $\delta=1$, is excluded because of the gauge invariance condition, $D_{A}=D_{A}^{i}G_{i}$~\cite{WessBagger}. 

In Sec.~\ref{sec:SUSYbreaking}, a model of SUSY breaking is introduced. This model is extended to include a scalar field (such as modulus or inflaton), and the direct gravitino production rate is studied in Sec.~\ref{sec:combined}. In particular, the conditions (I) and (II) obtained in the previous section are assumed to be satisfied. It will be explicitly shown that the rate is reduced compared to the F-term SUSY-breaking models, as the D-term SUSY-breaking effect increases. The cosmological implications are discussed in Sec.~\ref{sec:cosmology}.

%%%%%%%%%%%
\subsection{D-term SUSY-Breaking Model}
\label{sec:SUSYbreaking}

Let us consider a model of the D-term SUSY breaking explored in Ref.~\cite{Gregoire:2005jr}.
The model has a $U(1)$ gauge superfield and six chiral superfields, $z_{-1}, z_{1}, z_{-1/N}, z_{1/N}, z_{0}$ and $z'_{0}$, with their $U(1)$ charges denoted by the subscripts.
The K\"{a}hler potential and the gauge kinetic function are minimal, while the superpotential is
\begin{eqnarray}
W(z^i) = \lambda_1 z_0 \left( m^{1-N} z_1 z_{-1/N}^N-m^{2} \right) + \lambda_2 m z_1 z_{-1}
+ \lambda_3 z'_0 z_{1/N} z_{-1/N},
\label{eq:modelW}
\end{eqnarray}
where $m$ determines the SUSY-breaking scale.  
For $\lambda_{3}\gg \lambda_{1} \gg \lambda_{2} \gg g$, 
the field VEVs in the global SUSY limit are given by~\cite{Gregoire:2005jr}
\begin{eqnarray}
&& 
\braket{z_{-1}}= \braket{z_{1/N}}= \braket{z_{0}}=\braket{z'_{0}}=0,
\label{eq:vev}
\\
&&
|\braket{z_{-1/N}}|\simeq m\left( N^{3}\hat{\lambda}_{2}^{2} \right) ^{\frac{1}{2(N+2)}}, \label{eq:vev2}
 \\
&& \braket{z_{1}}\simeq m^{N+1}\braket{z_{-1/N}}^{-N}, \label{eq:vev3}
\end{eqnarray}
where $\hat{\lambda}_{2} \equiv \lambda_2/g$. 
At the vacuum, the scalar potential in the global SUSY is~\cite{Gregoire:2005jr}
\begin{align}
V \simeq g^{2} m^{4} \left( N^{3}\hat{\lambda}_{2}^{2} \right)^{\frac{2}{N+2}} \left( \frac{1}{N^{3}}+\frac{1}{2 N^{2}} \right).\label{Vapprox}
\end{align}
Here, $1/N^{3}$ and $1/2N^{2}$ in the parenthesis correspond to the F-term and D-term potentials, respectively. 
Thus, the model involves the D-term SUSY-breaking effect as well as that of the F term, and the fraction of the D-term SUSY breaking is approximately given by $\delta\simeq N/(N+2)$. The D-term contributions increase as $N$ becomes larger.

In the following numerical calculations, we use more precise values of the field VEVs, which are obtained by following the analysis in Appendix~\ref{sec:potential}, and the mass spectrum, which is obtained using the mass matrices listed in Appendix~\ref{sec:mass}.
The nonzero VEVs and mass spectra are shown in Table~\ref{tab:mass_spectrum}
for some model points, as well as the SUSY-breaking fraction $\delta$. The numerical results in Table~\ref{tab:mass_spectrum} agree with the approximate ones within $\sim 20\%$.

As shown in the table, all the fields in the SUSY-breaking sector have masses scaled by $m$, which is related to the gravitino mass $m_{3/2}$ by Eq.~\eqref{eq:m_and_m32}, or approximately 
\begin{align}
m \simeq 6^{\frac{1}{4}}N^{\frac{1}{2}}g^{-\frac{1}{2}}\sqrt{m_{3/2}M_{P}},
\end{align}
for large $N$, where we have used Eqs.~\eqref{V2}, \eqref{eq:vev2}, and \eqref{eq:vev3}.
Note that there is no light scalar field (the so-called Polonyi field) in the SUSY-breaking sector with a mass on the order of the gravitino mass.

%%%%%%%%%%%%%
\begin{table}[t!]
\begin{center}
\caption{
The fractions of the D-term SUSY breaking $(\delta )$, the field VEVs, and the mass spectra for $N=1, 10$, and 100. Here, two sets of the parameters, P1 and P2, are chosen. 
See Appendixes~\ref{sec:potential} and \ref{sec:mass} for details of the analysis.
The field VEVs and masses are written in units of $m$, except for the gravitino mass $m_{3/2}$.
}
\begin{tabular}{|c|l|c|c|cc|l|c|}
\hline
& $\lambda_{3}, \lambda_{1},$
&  
& 
& &
& \multicolumn{2}{|c|}{mass spectrum}
\\ \cline{7-8}
& $\lambda_{2}, g$ 
& $N$
& $\delta$
& $\vev{z_1}$ & $\vev{z_{-1/N}}$
& \multicolumn{1}{|c|}{\cancel{SUSY} sector}
& $m_{3/2}$
\\ \hline\hline
P1 
& $1, 10^{-1},$ 
& $1$
& $0.31$
& $0.46$ & $2.2$
& $3.7\times 10^{-3} - 2.2$ 
& $3.2\times 10^{-3}\, m^2/M_P $
\\
& $10^{-2}, 10^{-3}$
& $10$
& $0.83$
& $8.3\times 10^{-3}$ & $1.6$
& $5.6\times 10^{-4} - 12$ 
& $1.2\times 10^{-4}\, m^2 /M_P$
\\
&
& $100$
& $0.97$
& $1.4 \times 10^{-4}$ & $1.1$
& $1.3\times 10^{-4} - 7.0\times 10^{2}$ 
& $5.0 \times 10^{-6}\, m^2/M_P$
\\ \hline\hline
P2
& $1, 4^{-1}$,  
& $1$
& $0.27$
& $0.61$ & $1.6$
& $4.0\times 10^{-2} - 1.6$ 
& $2.6\times 10^{-2} m^2 / M_P$
\\
& $4^{-2}, 4^{-3}$
& $10$
& $0.83$
& $1.8 \times 10^{-2}$ & $1.5$
& $8.1 \times 10^{-3}  - 14$ 
& $1.6 \times 10^{-3} m^2 / M_P$
\\
&
& $100$
& $0.98$
& $2.9\times 10^{-4}$ & $1.1$
& $1.7\times 10^{-3} - 8.5 \times 10^{2}$ 
& $7.6 \times 10^{-5} m^2 / M_P$
\\ \hline
\end{tabular}

\label{tab:mass_spectrum}
\end{center}
\end{table}

%%%%%%%%%%%%%
\subsection{Gravitino Production Rate}
\label{sec:combined}

The model in the previous subsection is extended to include the scalar $\phi$. For simplicity, it is assumed that the K\"{a}hler potential and gauge kinetic function are minimal,
\begin{align}
K=
|\phi |^{2}+\left|z_{ 0} \right|^{2}+\left|z'_{0 } \right|^{2}+
\left|z_{1 } \right|^{2}+\left|z_{-1 } \right|^{2}+
\left|z_{1/N } \right|^{2}+\left|z_{-1/N } \right|^{2},~~~
h_{U(1)}=1.
\label{eq:modelKh}
\end{align}
It is also assumed that the mixings between $\phi$ and $z^i$ are absent in the superpotential, $W^{(n)}_{\rm mix}(\phi, z^i)=0$, and the total superpotential is given by
\begin{align}
W=W(z^i) +W(\phi),
\end{align}
with $W(z^i)$ given by Eq.~\eqref{eq:modelW}.
It is found that conditions (I) and (II) discussed in the previous section are satisfied. 

Let us now calculate the partial decay rate of the scalar $\phi$ into gravitinos, or the effective coupling constants in Eqs.~\eqref{GeffR} and \eqref{GeffI}, following the discussion in the previous section. They are determined by $G_\phi$ and the mixings between the scalar and the SUSY-breaking fields.
In the present model,
from Eq.~\eqref{Gphi}, $G_\phi$ becomes
\begin{align}
G_{\phi} \simeq - \frac{G_{z^{i}} \nabla_{\bar{\phi}}G_{\bar{z}^{i}}}{\nabla_{\bar{\phi}}G_{\bar{\phi}}},
\label{eq:Gphi_approx}
\end{align}
and the mixings in Eqs.~\eqref{eq:mass-matrix-1}, \eqref{eq:mass-matrix-2}, \eqref{Mphizbar}, and \eqref{Mphiz} are
\begin{align}
&\tilde{V}_{\phi \bar{z}^{i}} = V_{\phi \bar{z}^{i}} \simeq 
e^{G} \left( \nabla_{\phi}G_{\phi}\nabla_{\bar{z}^{i}}G_{\bar{\phi}} 
+ \nabla_{\phi}G_{z^{j}}\nabla_{\bar{z}^{j}}G_{\bar{z}^{i}} \right),
\label{eq:mix_approx}\\
&\tilde{V}_{\phi z^{i}} = V_{\phi z^{i}} \simeq 0.
\end{align}
In Eqs.~\eqref{eq:Gphi_approx} and \eqref{eq:mix_approx}, the factors $e^G$, $\nabla_\phi G_\phi$, and $\nabla_{z^i}G_{z^j}$ are of zeroth order in the perturbation, and are given by
\begin{align}
e^{G/2} = m_{3/2}  \simeq \vev{W},~~~
\nabla_\phi G_\phi \simeq \frac{m_\phi}{m_{3/2}},~~~
\nabla_{z^i}G_{z^j} = \frac{W_{z^i z^j}}{W} - \frac{W_{z^i}}{W} \frac{W_{z^j}}{W}.
\label{eq:nablaGzeroth}
\end{align}
On the other hand, the factor $\nabla_{z^i} G_{\phi} = \nabla_{\phi} G_{z^i}$ is suppressed by the Planck scale. Due to the absence of the $\phi - z^i$ mixings in $K$ and $W$
in the present model, it is given by
\begin{align}
\nabla_{\phi} G_{z^i} =
-
\frac{W_\phi}{W}
\frac{W_{z^i}}{W}
=
(K_\phi-G_\phi)\frac{W_{z^i}}{W}
\simeq K_\phi\frac{W_{z^i}}{W}.
\label{eq:Gzi_approx}
\end{align}
Here, in the last equality, we have used $G_\phi \simeq K_\phi \times {\cal O}(m_{3/2} / m_\phi) \ll K_\phi$, which can be shown by using Eq.~\eqref{eq:Gphi_approx},  $\nabla_\phi G_\phi \simeq m_\phi/m_{3/2}$,  and $G_{z^i}\simeq W_{z^i}/W \lesssim 1$.
See also Appendix~\ref{app:F-term}.
The remaining task is to evaluate the F-term VEVs of the SUSY-breaking fields, $W_{z^i}$, in the global SUSY limit. From Eqs.~\eqref{eq:modelW} -- \eqref{eq:vev3}, the finite F terms are
\begin{align}
W_{z_0} = \lambda_{1}\left(m^{1-N}\vev{z_{1}}\vev{z_{-1/N}}^{N}-m^{2}\right),~~~
W_{z_{-1}} = \lambda_2 m \vev{z_1}.
\label{eq:Wderiv}
\end{align}
From Eqs.~\eqref{eq:vev} -- \eqref{eq:vev3}, it is found that $W_{z_{-1}}$ dominates the F-term potential.\footnote{Although $W_{z_0}$ seems to vanish when the VEVs in Eqs.~\eqref{eq:vev} -- \eqref{eq:vev3} are naively applied, it is finite as long as the D term is nonzero, because the right-hand side of $W_{z_0}$ is analytically related to the D-term VEVs via the stationary condition of the scalar potential [Eq.~\eqref{eq:appV}]. 
The gaugino masses of the SUSY SM can be given by  $W_{z_0}$, or by the anomaly mediation.
}
The other F terms are zero in the global SUSY. 
 From Eqs.~\eqref{eq:Gzi_approx} and \eqref{eq:Wderiv}, it is found that $\nabla_{\phi} G_{z^i}$ 
is dominated by $\nabla_{\phi}G_{z_{-1}}$ as
\begin{align}
\nabla_{\phi}G_{z_{-1}} \simeq \frac{\lambda_{2} m}{m_{3/2}} \vev{z_{1}}\vev{\bar{\phi}},
\end{align}
where $\vev{W} \simeq m_{3/2}$ is used. 
Thus, the couplings are dominated by $z^i = z^j = z_{-1}$.
Putting the above equations altogether, one obtains
\begin{align}
\left| \mathcal{G}^{\text{(eff)}}_{\Phi_{R,I}} \right|^{2}
\simeq \left| 
\left( 1- m^{2}_{\phi} \left [ \left( m^{2}_{\phi}-m^{2}_{z} \right)^{-1} \right ]^{z_{-1}\bar{z}_{-1}} \right)
\frac{m_{3/2}}{m_\phi} G_{z_{-1}} \nabla_{\bar{\phi}} G_{\bar{z}_{-1}}
\right|^2.
\end{align}
For the purpose of the cosmological application in the next subsection, we concentrate on the case in which the scalar $\phi$ is lighter than the fields of the SUSY-breaking sector $z^{i}$. Then, the effective coupling constants are reduced to
\begin{align}
\left| \mathcal{G}^{\text{(eff)}}_{\Phi_{R,I}} \right|^{2}
\simeq
\left| 
\frac{m_{3/2}}{m_\phi} G_{\bar{z}_{-1}} \nabla_{\phi} G_{z_{-1}}
\right|^2
=
\left| 
\frac{\lambda_{2}^2 m^2}{m_{3/2} m_\phi} \vev{z_{1}}^2\vev{\bar{\phi}}
\right|^2
\simeq
\frac{36}{N^2} \frac{m_{3/2}^2}{m_\phi^2}\vev{\phi}^2.
\label{eq:modelGeff}
\end{align}
In the last equality we have used  Eqs.~\eqref{eq:vev2} and \eqref{eq:vev3}, and $N\gg 1$.  Here, the effective couplings become insensitive to the model parameters of the SUSY-breaking sector except for $N$.

\begin{figure}
  \begin{center}
   \includegraphics[width=85mm]{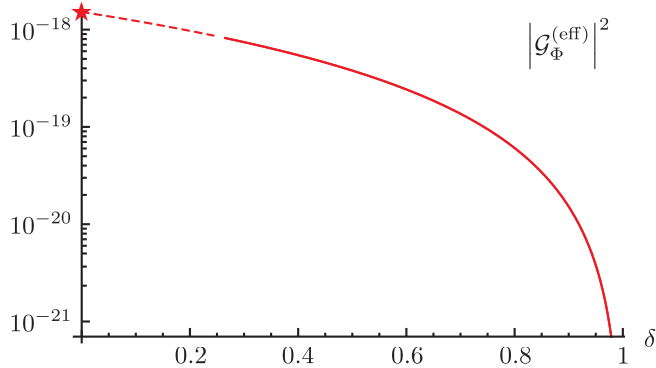}
  \end{center}
  \caption{Dependence of the effective coupling squared $\left | \mathcal{G}^{\text{(eff)}}_{\Phi_{R}} \right |^{2} \simeq \left | \mathcal{G}^{\text{(eff)}}_{\Phi_{I}} \right |^{2} \equiv \left | \mathcal{G}^{\text{(eff)}}_{\Phi} \right |^{2}$ on the fraction of D-term SUSY breaking $\delta$.  $m_{3/2}=10^3\GeV$, $m_{\phi}=10^{6}\GeV$, and $\vev{\phi}=10^{12}\GeV$ are used.  The lines for parameter sets P1 and P2 in Table~\ref{tab:mass_spectrum} coincide.
The fraction $\delta$ has been extrapolated to $N<1$, which is drawn by the dashed line.  The star mark represents the case of pure F-term SUSY breaking.}
  \label{fig:Geff}
\end{figure}

In Fig.~\ref{fig:Geff}, we show the effective coupling constants as a function of the fraction of the D-term SUSY breaking $\delta$, which is varied by changing the $N$ of the model. Here, we have used the full formulas in Eqs.~\eqref{GeffR} and \eqref{GeffI} with the numerically obtained field VEVs.\footnote{We have also used Eqs.~\eqref{eq:nablaGzeroth} and \eqref{eq:Gzi_approx} to evaluate the derivatives of $W(\phi)$.} The mass of the scalar $\phi$
is set to be smaller than the masses of the fields in the SUSY-breaking sector.
For the sake of comparison, we have also shown, by the star sign in the figure, the case where the SUSY is broken only by a single F-term VEV, assuming the minimal K\"{a}hler potential:
\begin{align}
\left| \mathcal{G}^{\text{(eff)}}_{\Phi_{R,I}} \right|^{2}
\simeq
\left|
\frac{G_z \nabla_{\bar{\phi}} G_{\bar{z}}}{\nabla_{\bar{\phi}} G_{\bar{\phi}}} 
\right|^2
\simeq
\left| \frac{3m_{3/2}}{m_\phi} \vev{\phi} \right|^2
\quad (\text{pure F term}),
\end{align}
where we have used $|G_z|\simeq \sqrt{3}$.
It is found that the effective coupling constants diminish as the D-term SUSY-breaking contribution increases.

%%%%%%%%%%%%%
%%%%%%%%%%%%%

%%%%%%%%%%%%%
\subsection{Cosmological Implications}
\label{sec:cosmology}

Finally, we estimate the total gravitino abundance.
The coherent oscillation of the scalar $\phi$ is assumed to dominate the energy density of the universe and then decay to create the thermal bath.
There are two types of gravitino production:
one is thermal production, where the gravitinos are produced through scattering processes in the thermal bath, and the other is direct production, which is the main subject of this paper.
In the following, we focus on the parameter region where the scalar is lighter than the fields in the SUSY-breaking sector except for the gravitino. Otherwise, the anomaly-induced gravitino production and the direct production of the SUSY-breaking fields which are mentioned in the last section can spoil the success of the standard cosmology.

First of all, let us consider thermal production of the gravitino.
The gravitino yield is defined as $Y_{3/2}=n_{3/2}/s$, where $n_{3/2}$ is the number density of the gravitino, and $s$ is the entropy density. The thermal gravitino yield is approximately given by~\cite{Kawasaki:2008qe}
\begin{align}
Y_{3/2}^{\text{thermal}}\simeq 
\left( 1.3\times 10^{-14}+8.8\times 10^{-15} \left( \frac{m_{1/2}}{m_{3/2}}\right)^{2} \right)
\left( \frac{T_{R}}{10^{8}\GeV} \right),
\label{Ythermal}
\end{align}
where $T_{R}=(\pi^{2}g_{*}(T_{R})/90)^{-1/4}\sqrt{\Gamma_\phi}$ is the reheating temperature after the $\phi$ decay, with the total decay rate $\Gamma_\phi$, and $m_{1/2}$ is the unified gaugino mass at the grand unified theory (GUT) scale.\footnote{The definition of $T_R$ here is different from the one in Ref.~\cite{Kawasaki:2008qe}.
Logarithmic corrections are omitted in Eq.~\eqref{Ythermal}.}

Next, let us consider the direct production of gravitinos.
The direct production of a pair of gravitinos provides
\begin{align}
Y_{3/2}^{\text{decay}} 
&= \frac{3T_{R}}{4m_{\phi}} \times 2B_{3/2} 
= \frac{1}{192\pi} \left( \frac{90}{\pi^{2} g_{*}(T_{R})} \right)^{\frac{1}{2}} \frac{m_{\phi}^{4} \left| \mathcal{G}_\Phi^{\text{(eff)}} \right|^{2}}{m_{3/2}^{2}T_{R}}.
\label{eq:yield}
\end{align}
Here, $m_{\Phi_{R}} \simeq m_{\Phi_{I}} \equiv m_{\phi}$ and $\left| \mathcal{G}^{\text{(eff)}}_{\Phi_{R}} \right|^{2} \simeq \left| \mathcal{G}^{\text{(eff)}}_{\Phi_{I}} \right|^{2} \equiv \left| \mathcal{G}_\Phi^{\text{(eff)}} \right|^{2}$ are used. 
 Substituting Eq.~\eqref{eq:modelGeff} into Eq.~\eqref{eq:yield}, the yield is reduced to be
\begin{align}
Y_{3/2}^{\text{decay}} 
&\simeq
\frac{1}{192\pi} \left( \frac{90}{\pi^{2} g_{*}(T_{R})} \right)^{\frac{1}{2}} 
\frac{1}{T_R} m_\phi^2 \vev{\phi}^2
\left(\lambda_2^4 \frac{\vev{z_1}^4 m^4}{m_{3/2}^4} \right) \nonumber\\
&\simeq
8.2 \times 10^{-15} \times
\frac{1}{N^2} \left( \frac{\vev{\phi}}{10^{12}\GeV} \right)^{2} \left( \frac{m_{\phi}}{10^{12}\GeV} \right)^{2} \left( \frac{10^{5}\GeV}{T_{R}} \right),
\label{Ydecay}
\end{align}
for large $N$, where $g_{*}(T_{R})=228.75$ is used. It is found that the yield is proportional to $m_\phi^2, \vev{\phi}^2$, and $T_R^{-1}$.  In particular, 
it becomes independent of $m_{3/2}$ because $\vev{z_1}\propto m \propto \sqrt{m_{3/2}}$. It is stressed that the abundance decreases as $N$ increases; namely, the D-term contributions dominate (See also Fig. \ref{fig:Geff}). 

\begin{figure}[thb]
  \begin{center}
   \includegraphics[width=85mm]{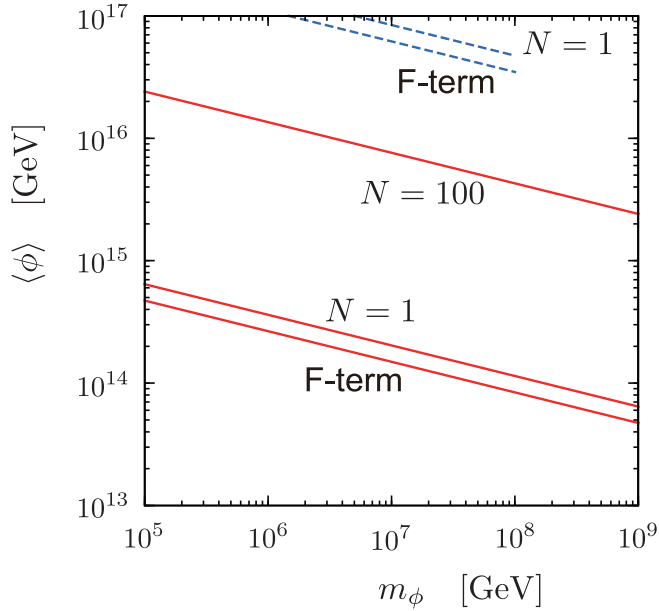}
  \end{center}
  \caption{Contours of the cosmological constraints on the $m_{\phi}$ -- $\vev{\phi}$ plane from the thermal and direct production of gravitinos.  The upper-right regions are excluded.
Several values of the D-term SUSY-breaking contributions are taken: $\delta=0$ (the pure F-term; see Fig.~\ref{fig:Geff} and the text), $0.27\ (N=1)$, and $0.98\ (N=100)$. 
The dashed lines correspond to $m_{3/2}=10$GeV, and the solid ones correspond to $m_{3/2}=1$TeV.
The former lines end in the figure because $m_{\phi}$ is restricted to be smaller than the smallest mass of the fields in the SUSY-breaking sector.
The reheating temperature is provided by $\Gamma_\phi=m_{\phi}^{3}/8\pi$. 
The thermal abundance is subdominant in this figure.
 }
  \label{fig:APR_typ}
\end{figure}

\begin{figure}
  \begin{center}
   \includegraphics[width=85mm]{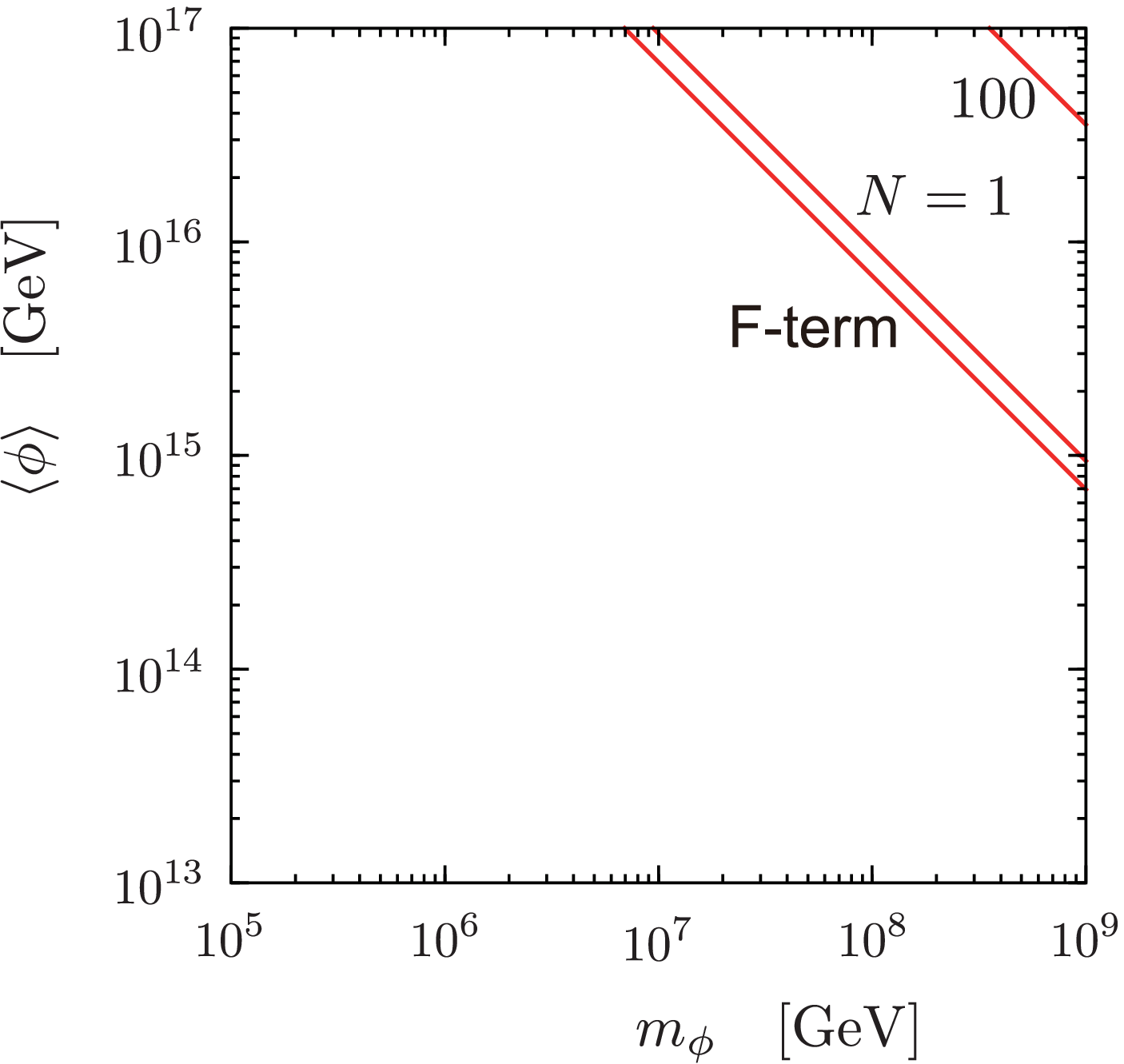}
  \end{center}
  \caption{Same as Fig.~\ref{fig:APR_typ}, except that the reheating temperature is optimized such that the total gravitino abundance is minimized. The bound on $m_\phi$ and $\vev{\phi}$ is considered to be most conservative. The gravitino mass is set to be $m_{3/2}=1\TeV$. There are no bounds in the case of $m_{3/2}=10$GeV in this parameter space.
 }
  \label{fig:APR_opt}
\end{figure}

The cosmological constraint on the gravitino abundance depends on the gravitino mass. 
When the gravitino is unstable, the most stringent bound comes from BBN.
The precise bound depends on details of the mass spectrum including the SUSY SM.
Here, we adopt the constraint for the model point in Case 2 of Ref.~\cite{Kawasaki:2008qe},\footnote{This model point is excluded by the recent SUSY search at the LHC, but we adopt it just for illustration. The main conclusion does not depend much on the details of the mass spectrum.}
 where the bounds are given by $Y_{3/2}< 2\times 10^{-16}$ and $Y_{3/2}< 2\times 10^{-12}$
for the gravitino masses of $m_{3/2}=1 \TeV$ and $30 \TeV$, respectively.
Thus, from Eqs.~\eqref{Ythermal} and \eqref{Ydecay}, the allowed region of the reheating temperature is 
\begin{align}
3 \times 10^6\GeV \times \frac{1}{N^2}
\left( \frac{\vev{\phi}}{10^{12}\GeV} \right)^{2} 
\left( \frac{m_{\phi}}{10^{12}\GeV} \right)^{2} 
< &\ T_{R} < 2\times 10^{6}\GeV
& (m_{3/2}=1\TeV),\\
5 \times 10^2\GeV \times \frac{1}{N^2}
\left( \frac{\vev{\phi}}{10^{12}\GeV} \right)^{2} 
\left( \frac{m_{\phi}}{10^{12}\GeV} \right)^{2} 
< &\ T_{R} < 1\times 10^{10}\GeV
& (m_{3/2}=30\TeV),
\end{align}
where the upper bound comes from the thermal gravitino abundance, while the lower bound is from direct production. This is because the thermal abundance is proportional to $T_R$, while the abundance from the latter is to $1/T_R$. 
It is noticed that the window tends to close as the energy scale of the scalar model is higher. In order to avoid the cosmological bound, the D-term contributions must be enhanced by increasing $N$. For instance, the window opens only when $N \gg 1$ for $m_{3/2}=1\TeV$, $\vev{\phi} = 10^{16}\GeV$, and $m_\phi = 10^9\GeV$. 

When the gravitino is the lightest among the SUSY particles, the gravitino becomes stable. 
Since the massive gravitino contributes to the dark matter abundance, the gravitino abundance cannot exceed the cold dark matter abundance as
\begin{align}
m_{3/2} Y_{3/2} < \frac{\rho_c}{s} \Omega_{\rm DM} < 4.4 \times 10^{-10}\GeV,
\end{align}
where $\rho_c$ is the critical density and $\Omega_{\rm DM} h^2 < 0.12$ at $2\sigma$~\cite{Beringer:1900zz} is used. 
For the light stable gravitino, the bound becomes
\begin{align}
1.9 \GeV \times\frac{1}{N^2}\lrf{m_{3/2}}{1 \GeV}
\left( \frac{\vev{\phi}}{10^{12}\GeV} \right)^{2} 
\left( \frac{m_{\phi}}{10^{12}\GeV} \right)^{2} 
< &\ T_{R} <
1.4 \times 10^{7}\GeV \times \lrf{m_{3/2}}{1\GeV},
\end{align}
where the gaugino mass is taken to be 600 GeV at the GUT scale, which affects the upper bound via the thermal gravitino abundance.
Again, the lower bound becomes milder as the D-term contributions dominate.

Finally, let us show how D-term SUSY-breaking contributions enlarge the allowed parameter space spanned by $m_{\phi}$ and $\vev{\phi}$. 
In Figs.~\ref{fig:APR_typ} and \ref{fig:APR_opt}, contours of the cosmological bounds are drawn for various $\delta$ in the $m_{\phi}$ -- $\vev{\phi}$ plane. The region of lower $m_{\phi}$ and $\vev{\phi}$ is allowed. 
In the figures, $m_\phi$ is restricted to be smaller than the smallest mass of the fields in the SUSY-breaking sector, for simplicity. 
It can be seen that as the D-term SUSY-breaking contributions increase---i.e., as $\delta$ becomes close to unity---the allowed region becomes wider.

The bound depends on the reheating temperature as seen in Eqs.~\eqref{Ythermal} and \eqref{Ydecay}. 
In Fig.~\ref{fig:APR_typ}, the temperature is estimated by assuming that the total decay rate is governed by a Planck-suppressed operator, $\Gamma (\phi \rightarrow \text{all}) = m_{\phi}^{3}/8 \pi$.
The thermal production is subdominant in the figure.
On the other hand, the reheating temperature is taken to be most conservative in Fig.~\ref{fig:APR_opt}. Since the thermal gravitino abundance is proportional to $T_R$, and that from the direct production is to its inverse, the total gravitino abundance becomes the minimum value by tuning $T_R$. Therefore, the bound in Fig.~\ref{fig:APR_opt} is the most conservative one, and the model is refuted if it is excluded in Fig.~\ref{fig:APR_opt}, unless there is an entropy production.

%%%%%%%%%%%%%%%%%%%%%%%%%%%%%%%%%%%%%%%%%%%%%%%%%%%%%%%%%%%%%%%%%%%%%%%%%%%%%%%%%%%%%%%%
\section{Summary}\label{sec:conclusion}

In this paper, we studied scalar decay into gravitinos in the presence of D-term SUSY-breaking and multiple SUSY-breaking fields, taking into account the mixing between the scalar and the fields in the SUSY-breaking sector.
We obtained the general formulas of the effective coupling constants for gravitino pair production,  Eqs.~\eqref{GeffR} and \eqref{GeffI}.
It is found that the gravitino production is suppressed when the D-term VEV is sizable if conditions (I) and (II) in Sec.~\ref{sec:EffCoupling} are satisfied.

As an example, we applied the formulas to the D-term SUSY-breaking model in Ref.~\cite{Gregoire:2005jr} and showed that the D-term SUSY breaking suppresses the gravitino production.
Moreover, an interesting feature of the model is that all the fields in the SUSY-breaking sector have masses of ${\cal O}(\sqrt{M_p m_{3/2}})$.
This is very different from F-term SUSY-breaking models, 
in which the SUSY-breaking field tends to have a mass of $\mathcal{O}(m_{3/2})$, 
causing the Polonyi problem~\cite{Coughlan:1983ci}. 
This may be a generic feature of D-term SUSY-breaking models. (See also
a recent work, Ref.~\cite{Azeyanagi:2012pc}.)
These observations can open a new window for model building respecting cosmology.

%%%%%%%%%%%%%%%%%%%%%%%%%%%%%%%%%%%%%%%%%%%%%
\section*{Acknowledgments}
This work was supported by Grants-in-Aid for Scientific Research from
the Ministry of Education, Science, Sports, and Culture (MEXT), Japan,
Grants No. 23740172 (M.E.), No. 21740164 (K.H.), and No. 22244021 (K.H.).
The work of T.T. was supported by an Advanced Leading Graduate Course for Photon Science grant.
This work was supported by the World Premier International Research Center Initiative (WPI Initiative), MEXT, Japan.

%%%%%%%%%%%%%%%%%%%%%%%%%%%%%%%%%%%%%%%%%%%%%
\appendix
%%%%%%%%%%%%%%%%%%%%%%%%%%%%%%%%%%%%%%%%%%%%%
\section{Gravitino Production by F-term SUSY Breaking}\label{app:F-term}

In this appendix, direct pair gravitino production by scalar decay is revisited when the SUSY breaking is caused by the F-term VEV. The derivation of the decay rate explored in this paper is somewhat different from that in the previous one \cite{Endo:2006tf}. In particular, the SUSY-breaking sector is generalized, and the sources of the gravitational corrections are clarified. 
In this section, the previous result will be reproduced. Moreover, the contribution which was overlooked will be pointed out. 

It is assumed here that the SUSY-breaking sector is composed of a single chiral supermultiplet, $z$, and there is no extra gauge symmetry for the sake of comparison with the result in Ref.~\cite{Endo:2006tf}. Also, the kinetic terms are supposed to be canonicalized in the global SUSY limit. 
In contrast to D-term SUSY breaking, the SUSY-invariant mass of $z$ is at most comparable to the gravitino mass, $\nabla_z G_z \sim 1$, which was neglected in Ref. \cite{Endo:2006tf}. However, the scalar mass of the SUSY-breaking field, $m_z$, can be lifted by a higher-dimensional operator, $\delta K = - |z|^4/\Lambda^2$ with $\Lambda \ll M_P$. Under the assumptions, the effective couplings [Eqs.~\eqref{GeffR} and \eqref{GeffI}] become
\begin{align}
\left| \mathcal{G}^{\text{(eff)}}_{\Phi_{R,I}} \right|^{2}=
\left| G_{\phi}+ G_{z} \frac{1}{m^{2}_{\Phi_{R,I}} - m_z^{2}} \left( \tilde{V}_{\bar{z} \phi} \pm \tilde{V}_{\bar{z} \bar{\phi}} \right) \right|^{2},
\end{align}
where $G_\phi$ and the mixing terms are simplified to be
\begin{align}
G_{\phi} &\simeq
-\frac{1}{\nabla_{\bar{\phi}}G_{\bar{\phi}}} G_{z} \nabla_{\bar{\phi}}G_{\bar{z}}
+ g_{\phi \bar z} G_{z},
\\
\tilde{V}_{\phi \bar{z}} &= 
V_{\phi \bar{z}} - g_{\phi \bar{z}} V^{\rm (g)}_{\phi\bar{\phi}} 
\nonumber \\ 
&\simeq e^{G} \left( \nabla_{\phi}G_{\phi}\nabla_{\bar{z}}G_{\bar{\phi}} 
- g_{z \bar \phi}\nabla_{\phi}G_{\phi}\nabla_{\bar{z}}G_{\bar{z}} 
- R_{\phi \bar{z} z \bar{z}} G_{\bar{z}} G_{z} \right) 
- g_{\phi \bar{z}} V^{\rm (g)}_{\phi\bar{\phi}},
\\
\tilde{V}_{\phi {z}} &= 
V_{\phi z} - g_{z\bar{\phi}} V^{\rm (g)}_{\phi \phi}
\simeq - e^{G} G_{\bar{\phi}z z}G_{\phi\phi} G_{\bar{z}} - g_{z\bar{\phi}} V^{\rm (g)}_{\phi \phi}.
\end{align}
The mass of the scalar $\phi$ is 
\begin{align}
m^{2}_{\Phi_{R,I}} = V^{\rm (g)}_{\phi \bar{\phi}} \pm V^{\rm (g)}_{\phi \phi}.
\end{align}
In particular, $V^{\rm (g)}_{\phi \bar{\phi}}$  is dominated by the superpotential term, $V^{\rm (g)}_{\phi \bar{\phi}} \simeq |W_{\phi\phi}|^2 \equiv m_\phi^2$, which is related to the SUGRA term as $e^{G/2} \nabla_\phi G_\phi \simeq W_{\phi\phi}$. On the other hand, the last term $V^{\rm (g)}_{\phi \phi}$ is small, which originates, e.g., in the SUSY-breaking effect, $V^{\rm (g)}_{\phi \phi} (\equiv \Delta m_\phi^2) \sim m_{3/2} m_\phi$. 

The effective coupling is compared to that in Ref.~\cite{Endo:2006tf} by considering the following,
\begin{align}
\left| \mathcal{G}^{\text{(eff)}}_{\Phi} \right|^{2} \equiv &
\frac{1}{2} \left( 
\left| \mathcal{G}^{\text{(eff)}}_{\Phi_{R}} \right|^{2} + \left| \mathcal{G}^{\text{(eff)}}_{\Phi_{I}} \right|^{2} 
\right) 
\nonumber \\
=&
\left| 
\frac{m_z^2}{m^{2}_{\phi} - m_z^2} 
\left( 
\frac{G_{z} \nabla_{\bar{z}} G_{\bar{\phi}}}{\nabla_{\bar{\phi}} G_{\bar{\phi}}}  
- g_{\phi \bar{z}} G_{z} \right)
\right. \nonumber \\ &~~~~ \left. 
- G_{z} \frac{1}{m^{2}_{\phi} - m_z^{2}} 
e^{G} \left( g_{z \bar \phi}\nabla_{\phi}G_{\phi}\nabla_{\bar{z}}G_{\bar{z}} 
+ R_{\phi \bar{z} z \bar{z}} G_{\bar{z}} G_{z} \right) \right|^{2}
\nonumber \\ &
+ \left| G_{z} \frac{1}{m^{2}_{\phi} - m_z^{2}} 
 e^{G} G_{\phi \bar{z} \bar{z}}G_{\bar{\phi}\bar{\phi}} G_{z}  \right|^{2},
\label{eq:F-term-result}
\end{align}
where the corrections proportional to $\Delta m_\phi^2$ from $(m^{2}_{\phi_{R,I}} - m_z^{2})^{-1}$ partially cancel those from $V_{\bar{z}\bar{\phi}}$, and the remaining corrections are subleading and omitted. This result reproduces the gravitino production rate in the F-term SUSY-breaking models \cite{Endo:2006tf}.

Among the contributions in Eq.~\eqref{eq:F-term-result}, $\nabla_{z} G_{\phi}$ is represented in terms of $K$ and $W$ as
\begin{align}
\nabla_{z} G_{\phi} = K_{z \phi} + \frac{W_{z\phi}}{W} - \frac{W_z}{W} \frac{W_\phi}{W} - \Gamma_{z\phi}^i G_i\,,
\label{eq:NablaGphi}
\end{align}
where $W_\phi/W$ can be replaced by
\begin{align}
\frac{W_\phi}{W} = G_\phi - K_\phi \simeq - K_\phi+g_{\phi \bar{z}}G_{z},
\end{align}
 up to a correction of order $m_{3/2}/m_\phi$. This is because
\begin{align}
\frac{W_\phi}{W} \simeq& -\frac{m_{3/2}}{m_{\phi}} G_{z}\nabla_{\bar{\phi}}G_{\bar{z}}+g_{\phi \bar{z}}G_{z}-K_{\phi} \nonumber \\
\simeq & -\frac{m_{3/2}}{m_{\phi}}G_{z}\left(  K_{\bar z \bar \phi} + \frac{\bar W_{\bar z \bar \phi}}{\bar W} - \frac{\bar W_{\bar z}}{\bar W} \frac{\bar W_{\bar \phi}}{\bar W} - \Gamma_{\bar z \bar \phi}^{\bar i} G_{\bar i}\right)+g_{\phi \bar{z}}G_{z}-K_{\phi},
\end{align}
and hence
\begin{align}
\left( 1 + \mathcal{O}\left(\frac{m_{3/2}}{m_{\phi}}\right) \right) \frac{W_{\phi}}{W}=-K_{\phi}+g_{\phi \bar{z}}G_{z}+ \mathcal{O}\left(\frac{m_{3/2}}{m_{\phi}}\right)\left(  K_{\bar z \bar \phi} + \frac{\bar W_{\bar z \bar \phi}}{\bar W} - \Gamma_{\bar z \bar \phi}^{\bar i} G_{\bar i}\right).
\end{align}
 Thus, Eq.~\eqref{eq:NablaGphi} becomes
\begin{align}
\nabla_{z} G_{\phi}  \simeq K_{z \phi} + \frac{W_{z\phi}}{W} + \frac{W_{z}}{W} K_\phi-g_{\phi \bar{z}}G_{z}\frac{W_{z}}{W} - \Gamma_{z\phi}^z G_z.
\end{align}
Barring accidental cancellations, $\nabla_{z} G_{\phi} \sim \braket {\phi}$ is obtained especially from the third term on the right-hand side. In Eq.~\eqref{eq:F-term-result}, the terms in the first parenthesis of the rightmost side, which include the $\nabla_{\bar{z}} G_{\bar{\phi}}$ term, are found to be suppressed if the scalar mass, $m_\phi^2$, is much larger than that of the SUSY-breaking field, $m_z^2$. They originate in $G_\phi$ and the mixing $\tilde{V}_{\phi \bar{z}}$, which cancel each other, as pointed out in Ref.~\cite{Dine:2006ii}. However, this cancellation does not work for $m_\phi \lesssim m_z$, and the direct gravitino production takes place generally. 
On the other hand, the other terms in Eq.~\eqref{eq:F-term-result} depend on the higher-dimensional operators. 

The term proportional to  $g_{z \bar \phi}\nabla_{\phi}G_{\phi}\nabla_{\bar{z}}G_{\bar{z}}$  in Eq.~\eqref{eq:F-term-result} is missed in Ref.~\cite{Endo:2006tf} because $\nabla_{\bar{z}}G_{\bar{z}}$ was supposed to be zero. If it is finite in the global SUSY limit, the contribution can be relevant when $m_\phi$ is larger than $m_z$. 

%%%%%%%%%%%%%%%%%%%%%%%%%%%%%%%%%%%%%%%%%%%%%
\section{Potential analysis}\label{sec:potential}

In this appendix, the scalar potential of the model in Sec.~\ref{sec:example} is analyzed.
First of all, the parameters are taken to be real and positive by redefinition of the fields. 
In the global SUSY framework, the phases of fields are minimized, except for the would-be Nambu-Goldstone (NG) boson. Next, the VEVs of $z_{0}$ and $z'_{0}$ can be represented in terms of those of the other fields, since the potential is quadratic with respect to $z_{0}$ and $z'_{0}$.
Substituting them into the potential, it involves four real variables,
$z_1$, $z_{-1}$, $z_{1/N}$, and $z_{-1/N}$.
It has quartic and quadratic terms with respect to $z_{-1}$. The coefficient of the $z_{-1}^4$ term is positive, and the coefficient of the $z_{-1}^2$ term is also positive if the following condition is fulfilled,
\begin{align}
\frac{N^{2} \braket{z_{1}}^{2} \lambda_{2}^{2} }{N^{2} \braket{z_{1}}^{2} +\braket{z_{1/N}}^{2}+ \braket{z_{-1/N}}^{2}} - g^{2} \left( \braket{z_{1}}^{2} + \frac{1}{N} \left( \braket{z_{1/N}}^{2}- \braket{z_{-1/N}}^{2} \right) \right) > 0. \label{consistency1}
\end{align}
If this is fulfilled, the VEV of $z_{-1}$ vanishes.
Similarly,  the VEV of $z_{1/N}$ vanishes if the following condition is satisfied:
\begin{align}
\lambda_{3}^{2}\braket{z_{-1/N}}^{2}+\frac{g^{2}}{N}\left(\braket{ z_{1}}^{2}-\frac{1}{N}\braket{z_{-1/N}}^{2} \right) > 0. \label{consistency2}
\end{align}
The VEVs of $z_{0}$ and $z'_{0}$ also vanish if these conditions are satisfied.
Then the potential becomes a function of $z_{1}$ and $z_{-1/N}$:
\begin{align}
V=\lambda_{1}^{2}\left( m^{1-N}z_{1}z_{-1/N}^{N}-m^{2}\right)^{2}+\lambda_{2}^{2}m^{2}z_{1}^{2}+\frac{g^{2}}{2}\left( z_{1}^{2}-\frac{1}{N}z_{-1/N}^{2} \right)^{2}.
\label{eq:appV}
\end{align}
Finally, we obtain the VEVs of $z_1$ and $z_{-1/N}$ (in units of $m$) numerically by minimizing the potential, and confirm that the two consistency conditions [Eqs.~\eqref{consistency1} and \eqref{consistency2}] are satisfied.

Once the field VEVs are obtained, the SUSY-breaking scale can be written in terms of the gravitino mass by using Eqs.~\eqref{V} or \eqref{V2}:
\begin{align}
m = 3^{1/4} \left( \lambda_{1}^{2}(\hat{z}_{1}\hat{z}_{-1/N}^{N}-1)^{2}+\lambda_{2}^{2}\hat{z}_{1}^{2}+\frac{g^{2}}{2}\left( \hat{z}_{1}^{2}-\frac{1}{N}\hat{z}_{-1/N}^{2}\right)^{2} \right)^{-1/4}
\sqrt{m_{3/2}M_P},
\label{eq:m_and_m32}
\end{align}
where $\hat{z}_{1}$ and $\hat{z}_{-1/N}$ denote the absolute value of each VEV in units of $m$.

%%%%%%%%%%%%%%%%%%%%%%%%%%%%%%%%%%%%%%%%%%%%%
\section{Mass-matrix components at the vacuum}\label{sec:mass}
In this appendix, we explicitly write the mass matrix  of the model used in Sec.~\ref{sec:example} at the vacuum in the global SUSY.
In the following, all the parameters and VEVs have been transformed to real and positive variables.
In this appendix, we take $m=1$, and use a shorthand notation for field derivatives like $V^{\text{(g)}}_{i,\overline{j}}=V^{\text{(g)}}_{z^i\bar z^{j}}$.
\subsection{Mass-matrix components of the scalars of the SUSY-breaking sector}
The scalar mass matrix can be written in a block diagonal form:
\begin{align}
m^{2}_{z}=\left(
\begin{array}{ccc}
A & 0 & 0 \\
0 & B & 0 \\
0 & 0 & C 
\end{array}
\right),
\end{align}
where
\begin{align}
A=&\left( \begin{array}{cccc}
V^{\text{(g)}}_{1,\overline{1}} & V^{\text{(g)}}_{1,\overline{-1/N}} & V^{\text{(g)}}_{1,1} & V^{\text{(g)}}_{1,-1/N} \\
V^{\text{(g)}}_{-1/N,\overline{1}} & V^{\text{(g)}}_{-1/N,\overline{-1/N}} & V^{\text{(g)}}_{-1/N,1} & V^{\text{(g)}}_{-1/N,-1/N} \\
V^{\text{(g)}}_{\overline{1},\overline{1}} & V^{\text{(g)}}_{\overline{1},\overline{-1/N}} & V^{\text{(g)}}_{\overline{1},1} & V^{\text{(g)}}_{\overline{1},-1/N} \\
V^{\text{(g)}}_{\overline{-1/N},\overline{1}} & V^{\text{(g)}}_{\overline{-1/N},\overline{-1/N}} & V^{\text{(g)}}_{\overline{-1/N},1} & V^{\text{(g)}}_{\overline{-1/N},-1/N} 
\end{array}\right), \\
B=&\left( \begin{array}{cc}
b&0\\
0&b
\end{array}\right)
~~~\text{with}~~b=\left( \begin{array}{cc}
V^{\text{(g)}}_{-1,\overline{-1}}&V^{\text{(g)}}_{-1,\overline{0}}\\
V^{\text{(g)}}_{0,\overline{-1}}&V^{\text{(g)}}_{0,\overline{0}}
\end{array}\right), \\
C=&\left( \begin{array}{cccc}
c&0&0&0\\
0&c&0&0\\
0&0&c'&0\\
0&0&0&c'
\end{array}\right)
~~~\text{with}~~c=V^{\text{(g)}}_{1/N,\overline{1/N}}~~\text{and}~~c'=V^{\text{(g)}}_{0',\overline{0'}}.
\end{align}
The nonzero components are given below:
\begin{align}
V^{\text{(g)}}_{1,\overline{1}}=&\lambda_{2}^{2}+\lambda_{1}^{2}z_{-1/N}^{2N}+g^{2}\left( 2 z_{1}^{2}-\frac{1}{N}z_{-1/N}^{2}\right), \\
V^{\text{(g)}}_{-1/N,\overline{-1/N}}=&N^{2}\lambda_{1}^{2}z_{1}^{2}z_{-1/N}^{2(N-1)}-\frac{g^{2}}{N}\left(  z_{1}^{2}-\frac{2}{N}z_{-1/N}^{2}\right), \\
V^{\text{(g)}}_{1,\overline{-1/N}}=&N\lambda_{1}^{2}z_{1}z_{-1/N}^{2N-1}-\frac{g^{2}}{N}z_{1}z_{-1/N}, \\
V^{\text{(g)}}_{1,1}=&g^{2}z_{1}^{2}, \\
V^{\text{(g)}}_{-1/N,-1/N}=&N(N-1)\lambda_{1}^{2}z_{1}z_{-1/N}^{N-2}\left( z_{1}z_{-1/N}^{N}-1\right) +\frac{g^{2}}{N^{2}}z_{-1/N}^{2},\\ \displaybreak[2]
V^{\text{(g)}}_{1,-1/N}=& N \lambda_{1}^{2}z_{-1/N}^{N-1}\left( z_{1}z_{-1/N}^{N}-1 \right)-\frac{g^{2}}{N}z_{1}z_{-1/N}, \\ \displaybreak[2]
V^{\text{(g)}}_{-1,\overline{-1}}=&\lambda_{2}^{2}-g^{2}\left( z_{1}^{2}-\frac{2}{N}z_{-1/N}^{2} \right), \\ \displaybreak[2]
V^{\text{(g)}}_{0,\overline{0}}=&\lambda_{1}^{2}z_{-1/N}^{2N}+N^{2}\lambda_{1}^{2}z_{1}^{2}z_{-1/N}^{2(N-1)}, \\
V^{\text{(g)}}_{-1,\overline{0}}=&\lambda_{1}\lambda_{2}z_{-1/N}^{N},\\
V^{\text{(g)}}_{1/N,\overline{1/N}}=&\lambda_{3}^{2}z_{-1/N}^{2}+\frac{g^{2}}{N}\left( z_{1}^{2}-\frac{1}{N}z_{-1/N}^{2} \right), \\
V^{\text{(g)}}_{0',\overline{0'}}=&\lambda_{3}^{2}z_{-1/N}^{2}.
\end{align}
The mass matrix has one zero eigenvalue, which corresponds to  the would-be NG boson of the $U(1)$ gauge symmetry.

\subsection{Mass-matrix components of the fermions}
The nonzero components of the mass matrix components $M_{i,j}$ are listed below.
The subscript ``$U(1)$'' denotes the gaugino.
\begin{align}
M_{-1, 1}=&\lambda_{2}, \\
M_{1, 0}=&\lambda_{1}z_{-1/N}^{N}, \\
M_{-1/N, 0}=&N \lambda_{1}z_{1}z_{-1/N}^{N-1}, \\
M_{1, U(1)}=&-\sqrt{2}igz_{1}, \\
M_{-1/N, U(1)}=&\frac{\sqrt{2}}{N}igz_{-1/N}, \\
M_{1/N, 0'}=&\lambda_{3}z_{-1/N}.
\end{align}
The mass matrix has one zero eigenvalue, which corresponds to the would-be goldstino of the local SUSY.

%%%%%%%%%%%%%%%%%%%%%%%%%%%%%%%%%%%%%%%%%%%%

\end{document}